%% file: Zb.tex
\documentclass[preprint,5p,times,twocolumn]{elsarticle}

\usepackage{amssymb}
\usepackage{amsmath}
\usepackage{graphicx}
\graphicspath{{figures/}}
\usepackage{atlasphysics}
\usepackage{subfigure}
\usepackage{hyperref}
\usepackage{epsfig}
\usepackage{multirow}

\newcommand{\ppbar}{$p\bar{p}$}

\journal{Physics Letters B}

\begin{document}

\begin{frontmatter}

\title{\vspace{-0.5cm} \flushright{\normalsize{CERN-PH-EP-2011-133}} \\ \center{Measurement of the cross-section for $b$-jets produced in association with a $Z$ boson at $\sqrt{s}=7$~TeV with the ATLAS detector}}

\author{The ATLAS Collaboration}

\begin{abstract}
A measurement is presented of the inclusive cross-section for $b$-jet production in association with a $Z$ boson
in $pp$ collisions at a centre-of-mass energy of $\sqrt{s}=7$~TeV.
The analysis uses the data sample collected by the ATLAS experiment in 2010, corresponding
to an integrated luminosity of approximately 36~pb$^{-1}$.
The event selection requires a $Z$ boson decaying into high $\pt$ electrons or muons,
and at least one  $b$-jet, identified by its displaced vertex, with transverse momentum $\pt> 25~\GeV$
and rapidity $|y|<2.1$. After subtraction of background processes,
the yield is extracted from the vertex mass distribution of the candidate $b$-jets. The ratio of this 
cross-section to the inclusive $Z$ cross-section (the average number of $b$-jets per $Z$ event)
is also measured. Both results are found to be in good agreement with
perturbative QCD predictions at next-to-leading order.
\end{abstract}

\begin{keyword}
Standard Model, Z boson, b-jet, Cross-section
\end{keyword}

\end{frontmatter}

\section{Introduction}

The production of $Z$ bosons in association with jets at hadron colliders has
long been used as a testing ground for perturbative QCD (pQCD)
calculations. However, whilst substantial progress has been made in
understanding and modelling the production of inclusive jets in $Z$
events, the production of heavy flavour ($b$ or $c$) jets is less well
studied.  The production of one or more $b$-jets in association with a
$Z$ boson is a significant background to many important searches at
the LHC, such as the Standard Model Higgs search, SUSY searches and searches for 
other physics beyond the Standard Model. A measurement of $Z$ plus $b$-jets
production therefore directly improves the understanding of this process,
and consequently the ability to accurately model this background.

\begin{figure}[htp]
\begin{center}
\subfigure[]{\includegraphics[width=0.20\textwidth]{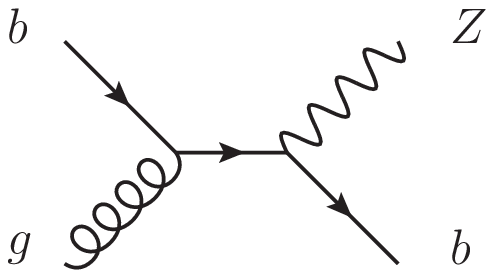}}\quad
\subfigure[]{\includegraphics[width=0.20\textwidth]{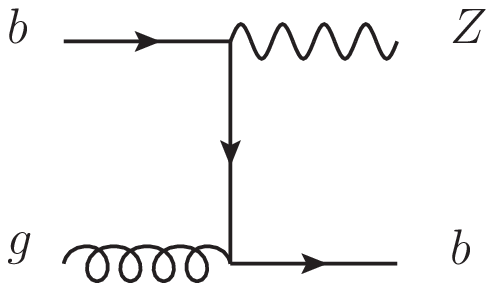}}\\
\subfigure[]{\includegraphics[width=0.20\textwidth]{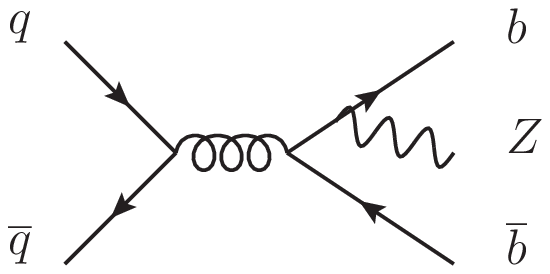}}\quad
\subfigure[]{\includegraphics[width=0.20\textwidth]{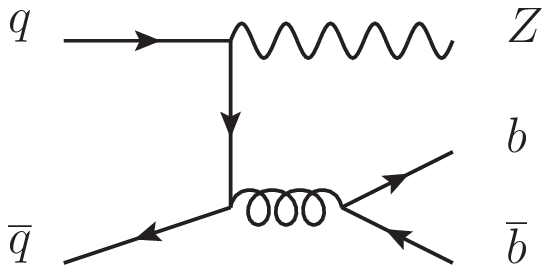}}
\caption{ Main diagrams for associated production of a $Z$ boson and one or more $b$-jets.}
\label{fig:diagrams}
\end{center}
\end{figure}

Figure~\ref{fig:diagrams} shows the main diagrams that contribute to $Z + b$ production.
The top two diagrams have an initial state $b$-quark, whereas in the bottom
two diagrams, a \bbbar\ pair is explicitly produced in the final state.
The ALPGEN \cite{alpgen} and SHERPA \cite{sherpa} Monte-Carlo (MC) generators implement leading order
(LO) calculations of this process using massive $b$-quarks. For the amplitudes with initial $b$-quarks, 
ALPGEN first creates a \bbbar\  pair from the distribution of gluons, and integrates over the whole 
phase space for these quarks. SHERPA, in the present implementation, draws a $b$-quark from a 
Parton Density Function (PDF) derived from the gluon distribution. These generators are interfaced to parton shower
 and hadronisation packages and provide direct comparison to the data.
In addition to the diagrams in Fig.~\ref{fig:diagrams}, it is also possible that the $Z$ boson 
and $b$-jets are produced in two different parton-parton collisions in the same proton-proton 
interaction. This process, referred to as multiple parton interaction (MPI), is included in the ALPGEN 
and SHERPA generators. 
In contrast, the MCFM programme \cite{MCFM} implements next-to-leading order (NLO) 
calculations~\cite{nlo4} using massless $b$-quarks, with initial $b$-quarks taken from a $b$-PDF. 
MCFM is not interfaced to parton shower/hadronisation packages, and does not include MPI. 
Calculations of the $Z + b$ process at NLO continue to be an active area of development~\cite{nlo5,nlo4,nlo3,nlo2,nlo1}.

Previous measurements of $Z + b$ production at lower centre-of-mass 
energy \ppbar\ collisions at the Fermilab Tevatron collider by the CDF and D0 collaborations
\cite{CDF-Zb,D0-Zb} are consistent with pQCD calculations.
In this Letter we present a measurement of the inclusive
cross-section for $b$-jet
production in association with a $Z$ boson, $\sigma_{b}$. 
The measurement is made at the particle-level, and is fully corrected for all
detector effects. A $b$-jet is defined here as a jet
which contains a $b$-hadron.
Here and in the following, $Z$ stands for both the $Z$ boson and 
virtual photon $\gamma^{*}$ contributions. The $Z$ boson is identified
by its decay into a pair of high transverse momentum, opposite sign
electrons (electron channel) or muons (muon channel), and the $Z$ and
$b$-jets are reconstructed
within the allowed fiducial coverage of the detector. The
cross-section $\sigma_{b}$ is quoted per lepton channel, within this
fiducial coverage. A closely related measurement has been performed, using very similar techniques, in the
$W + b$ final state \cite{Wbjets}.

\section{The ATLAS Detector}
\label{detector}

The ATLAS detector \cite{atlas-detector} consists of an inner tracking system surrounded by a thin
superconducting solenoid providing a 2\,T axial magnetic field, electromagnetic
and hadronic calorimeters and a muon spectrometer. The inner detector system
provides tracking information for charged particles in a pseudorapidity range
\abseta~$<2.5$\footnote{The azimuthal angle $\phi$ is measured around the
  beam axis and the	polar angle $\theta$ is the angle from the beam
  axis. The pseudorapidity is defined as $\eta = -\rm{ln~tan}(\theta/2)$. The
  distance $\Delta R$ in $\eta-\phi$ space is defined as $\Delta R = 
\sqrt{\Delta\phi^{2} + \Delta\eta^{2} }$.}. 
At small radii, high granularity silicon pixel and microstrip detectors allow for
the reconstruction of secondary decay vertices. The electromagnetic calorimeter uses lead absorbers and liquid argon as
the active material and covers the rapidity range
\abseta~$<3.2$, with high longitudinal and transverse granularity for
electromagnetic shower reconstruction. For electron detection the transition
region between the barrel and end-cap calorimeters, 
$1.37<$~\abseta~$<1.52$, is not considered in this analysis. 
The hadronic tile calorimeter is a steel/scintillating-tile detector that extends the
instrumented depth of the calorimeter to fully contain hadronic
particle showers. In the forward regions it is complemented by two
end-cap calorimeters using liquid argon as the active material and
copper or tungsten as the absorber material.
The muon spectrometer comprises three large air-core superconducting toroidal magnets
which provide a typical field integral of 3\,Tm. Three stations of chambers provide
precise tracking information in the range \abseta~$<2.7$, and triggers for high momentum
muons in the range \abseta~$<2.4$.
The transverse energy $E_{T}$ is defined to be $E \rm{sin} \theta$, where $E$ is
the energy associated with a calorimeter cell or 
energy cluster. Similarly, $p_{T}$ is the momentum component transverse to the beam line.

\section{Collision data and simulated samples}
\label{samples}
\subsection{Collision data}
The analysis presented here is performed on data from $pp$ collisions at a centre-of-mass
energy of 7\,TeV recorded by ATLAS in 2010 in stable beams periods and uses data selected for good
detector performance.
The events were selected online by requiring at least one electron or muon with high transverse
 momentum, \pt. The trigger thresholds evolved with time to keep up with the increasing instantaneous
luminosity delivered by the LHC. The highest thresholds applied in the last data taking period
were \et~$>15$ GeV for electrons and $\pt>13\GeV$ for muons.  The integrated luminosity after
beam, detector and data-quality requirements is $36.2$\,\ipb ($35.5$\,\ipb) for events
collected with the electron (muon) trigger, measured with a $\pm 3.4\%$ relative
 error \cite{lumipaper,lumiconfnote}.

\subsection{Simulated events}
\label{simevents}
The measurements will be compared to theoretical predictions of the Standard Model, using
Monte-Carlo samples of signal and background processes. The detector response to the generated events
is fully simulated with GEANT4~\cite{geant}.

Samples of signal events containing a $Z$ boson decaying into electrons or muons and at least one $b$-jet
have been simulated using the ALPGEN, SHERPA, and MCFM generators, using the CTEQ6.6 PDF set \cite{cteq66}.
All three generators include $Z/\gamma^*$ interference terms. The ALPGEN generator is interfaced
to HERWIG \cite{herwig1} for parton shower and fragmentation, and JIMMY for the underlying event simulation
 \cite{herwig4}.
For jets originating from the hadronisation of light quarks or gluons (hereafter referred
to as $light$-jets), the LO generator ALPGEN uses MLM matching \cite{mlm} to remove any double
counting of identical jets produced via the matrix element and parton shower, but this is not
available for $b$-jets in the present version.
Therefore events containing two $b$-quarks with $\Delta R < 0.4$
 ($\Delta R >0.4$) coming from the matrix element (parton shower) contribution are removed.
SHERPA uses the CKKW \cite{ckkw} matching for the same purpose.
The MCFM NLO generator lacks an interface to a parton shower and fragmentation package, hence to compare
with the data we apply correction factors describing the parton-to-particle correspondence, obtained from
particle-level LO simulations.
For all Monte-Carlo events, the cross-section is normalised by rescaling the inclusive 
$Z$ cross-section of the relevant generator to the NNLO cross-section \cite{Z-cross}.

The dominant background comes from
$Z$ + jets events, with the $Z$ decaying into electrons, muons or tau leptons, where one jet is a $light$
or $c$-jet which has been incorrectly tagged as a $b$-jet.
These events are simulated using the same generators as the signal.
Other background processes considered include \ttbar\ pair production simulated by MC@NLO \cite{mcnlo,mcnlo2},
 $W(\rightarrow l\nu)$ + jets simulated by PYTHIA \cite{pythia1}, $WW$/$WZ$/$ZZ$ simulated by ALPGEN,
and single-top production simulated by MC@NLO.
The cross-sections for these processes have been
normalised to the predictions of
\cite{ttbar-cross,ttbar-cross2} (approximate NNLO) for \ttbar\ pair
production, \cite{Z-cross} (NNLO) for $W(\rightarrow l\nu)$ + jets,
\cite{MCFM} (NLO) for $WW$/$WZ$/$ZZ$, and the MC@NLO value for single-top.

Events have been generated with the number of collision vertices drawn from a Poisson
distribution with an average of 2.0 vertices per event. Simulated events are then reweighted to match the observed vertex
distribution in the data.

\section{Reconstruction and selection of $Z+b$ candidates}
\label{selection}

Events are required to contain one primary vertex with at least three
high-quality charged tracks.
As the final state should contain a $Z$ boson, the selection of events closely follows the selection
criteria used by ATLAS for the inclusive $Z$ analysis \cite{WZbaseline}.
In the $e^{+} e^{-}$ channel, two opposite sign electron candidates are required with $\et>20\GeV$ and
$\abseta<2.47$. Electron candidates are reconstructed from a cluster of cells in the electromagnetic
calorimeter and a charged particle track in the inner detector. Criteria are applied on the longitudinal
and transverse shower shapes in the calorimeters and on the matching of the track with the cell cluster, requesting a \it{Medium} \rm \cite{WZbaseline} electron quality.
Similarly in the $\mu^{+} \mu^{-}$ channel, two opposite sign muons are required with $\pt>20\GeV$ and
$\abseta<2.4$. Muon candidates are reconstructed from a track in the muon spectrometer associated
with a track in the inner detector. To reject cosmic rays the track is required to be compatible with
coming from the primary vertex of the collision under study. In addition, an isolation criterion is applied requiring that
the summed $\pt$ of tracks in a cone $\Delta R  = 0.2$ around the muon
candidate be less than 10\% of the muon \pt.
For both channels, the invariant mass of the lepton pair is then required to be $76<m_{ll}<106$ GeV.
This window is chosen to be narrower than that used in the inclusive $Z$ analysis in order to reduce the
background from \ttbar\ events, and multi-jet events where jets are mis-identified as leptons,
either due to a high electromagnetic content in the shower, or mis-reconstruction.

Jets are reconstructed from clusters of electromagnetic and hadronic
calorimeter cells using an anti-$k_{t}$~\cite{AntiKt} algorithm with a resolution parameter of $0.4$.
A jet calibration procedure is applied which includes an energy offset which depends on the number
of primary vertices, in order to minimise the impact of additional
(pile-up) collisions \cite{jes}. The $b$-jets do not receive any
additional calibration, thus their difference
in jet energy scale with respect to $light$-jets (2.5\% in simulated events) is treated as a systematic uncertainty, discussed in Section~\ref{syst}.
To avoid double-counting electrons and muons as jets, jets with $\Delta R <0.5$ to either
of the leptons coming from the $Z$ pair are removed. Events are
selected requiring at least one jet with \pt~$>25$~GeV and $|y|<2.1$.

 A jet is considered as $b$-tagged if the SV0 algorithm \cite{sv0} reconstructs a secondary vertex from charged particle tracks contained within the jet and the decay length significance of this secondary vertex is greater than 5.85. This requirement provides 50\%
efficiency for tagging $b$-jets in simulated \ttbar\ events
\cite{btag-eff}. In addition, the invariant mass of the charged particle tracks
from which the secondary vertex is reconstructed -- the SV0-mass --
will be used to extract the $b$-jet fraction on a statistical basis. Table \ref{tab:cutflow} gives the number of data events selected by the consecutive stages of the analysis.
\begin{table}[t]
\begin{center}
\begin{tabular}{lccc}
\hline\hline
 \multicolumn{2}{c}{Electron channel} &
  \multicolumn{2}{c}{Muon channel} \\
\hline
 Criterion &  Events & Criterion & Events \\
\hline
2 selected electrons &  10558  & 2 selected muons  & 13691  \\
$Z$ mass window &  9230   & $Z$ mass window  &  12222 \\
$\geq1$ jet&   1597  &$\geq1$ jet& 1987   \\
$\geq1$ $b$-tag& 64  & $\geq1$ $b$-tag  & 67     \\
\hline
$= 1$ $b$-tag& 62  & $= 1$ $b$-tag  & 63     \\
$= 2$ $b$-tag& 1  & $= 2$ $b$-tag  & 4     \\
$= 3$ $b$-tag& 1  & $= 3$ $b$-tag  & 0     \\
\hline\hline
\end{tabular}
\caption{The number of events selected at various stages of the
  analysis event selection.}
\label{tab:cutflow}
\end{center}
\end{table}

\section{Background subtraction and signal yield determination}
\label{yield}
\subsection{Background estimation}
\begin{figure*}[t!]
\begin{center}
\subfigure{\includegraphics[width=0.49\textwidth]{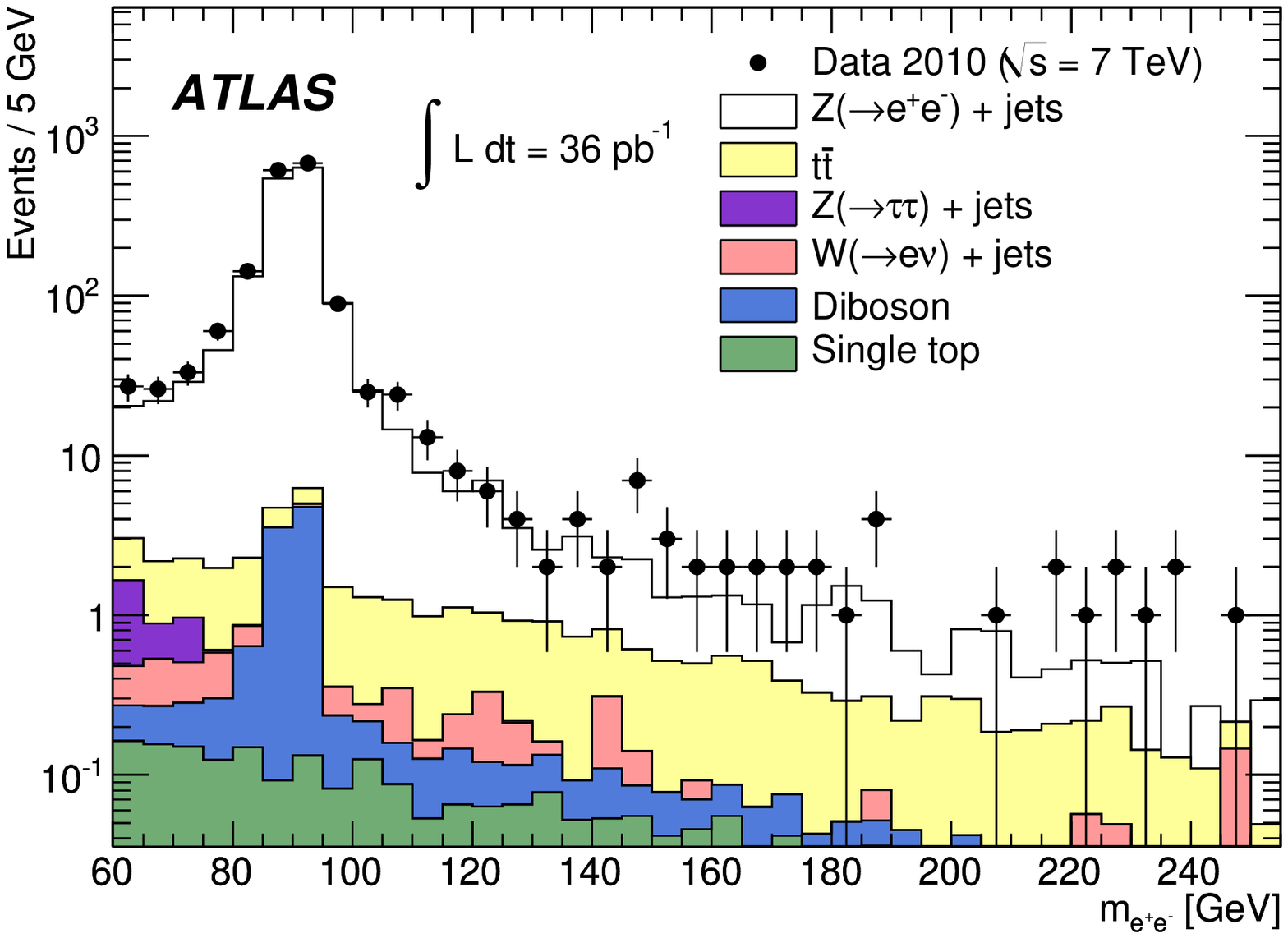}}
\subfigure{\includegraphics[width=0.49\textwidth]{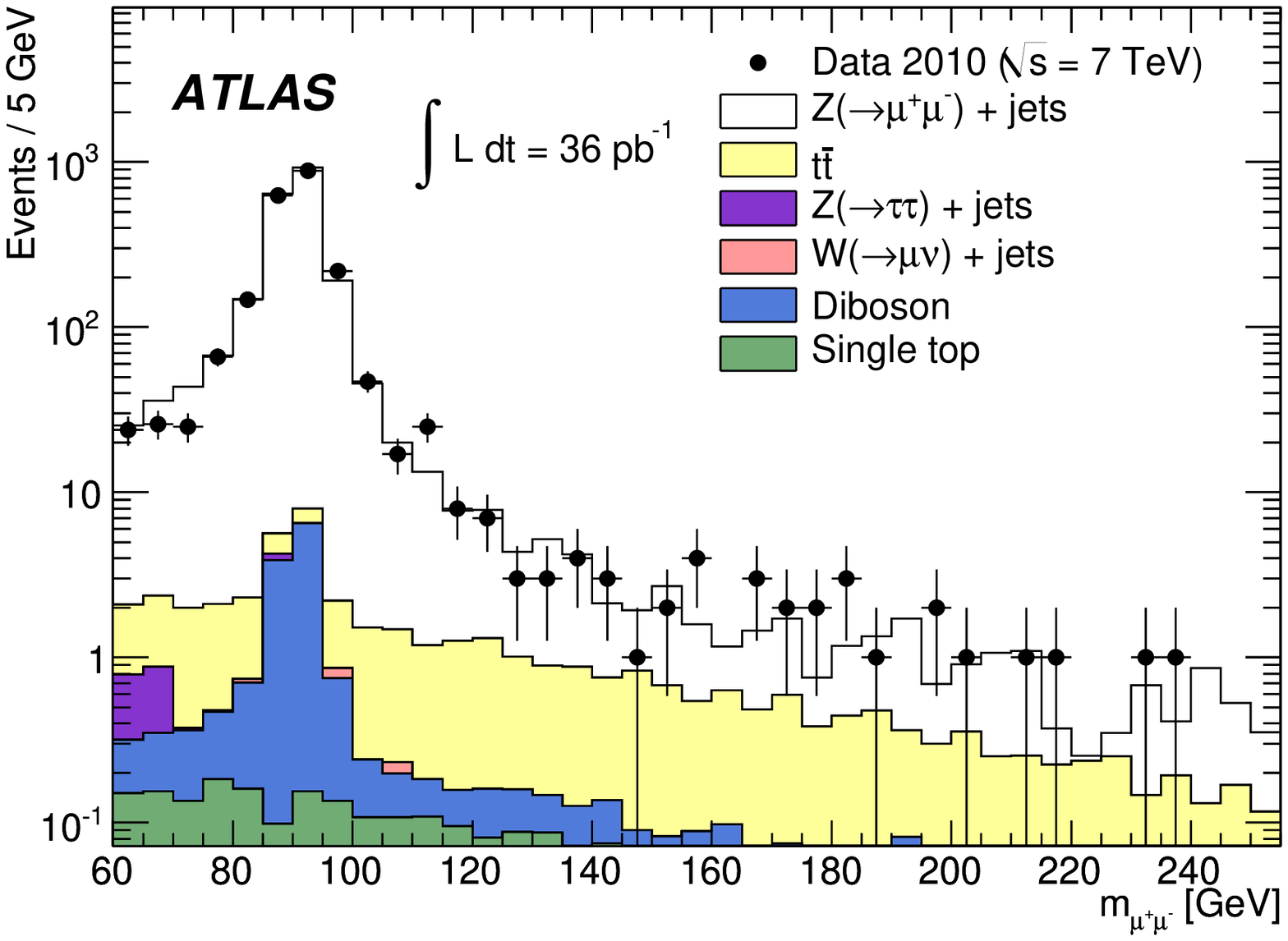}}
\subfigure{\includegraphics[width=0.49\textwidth]{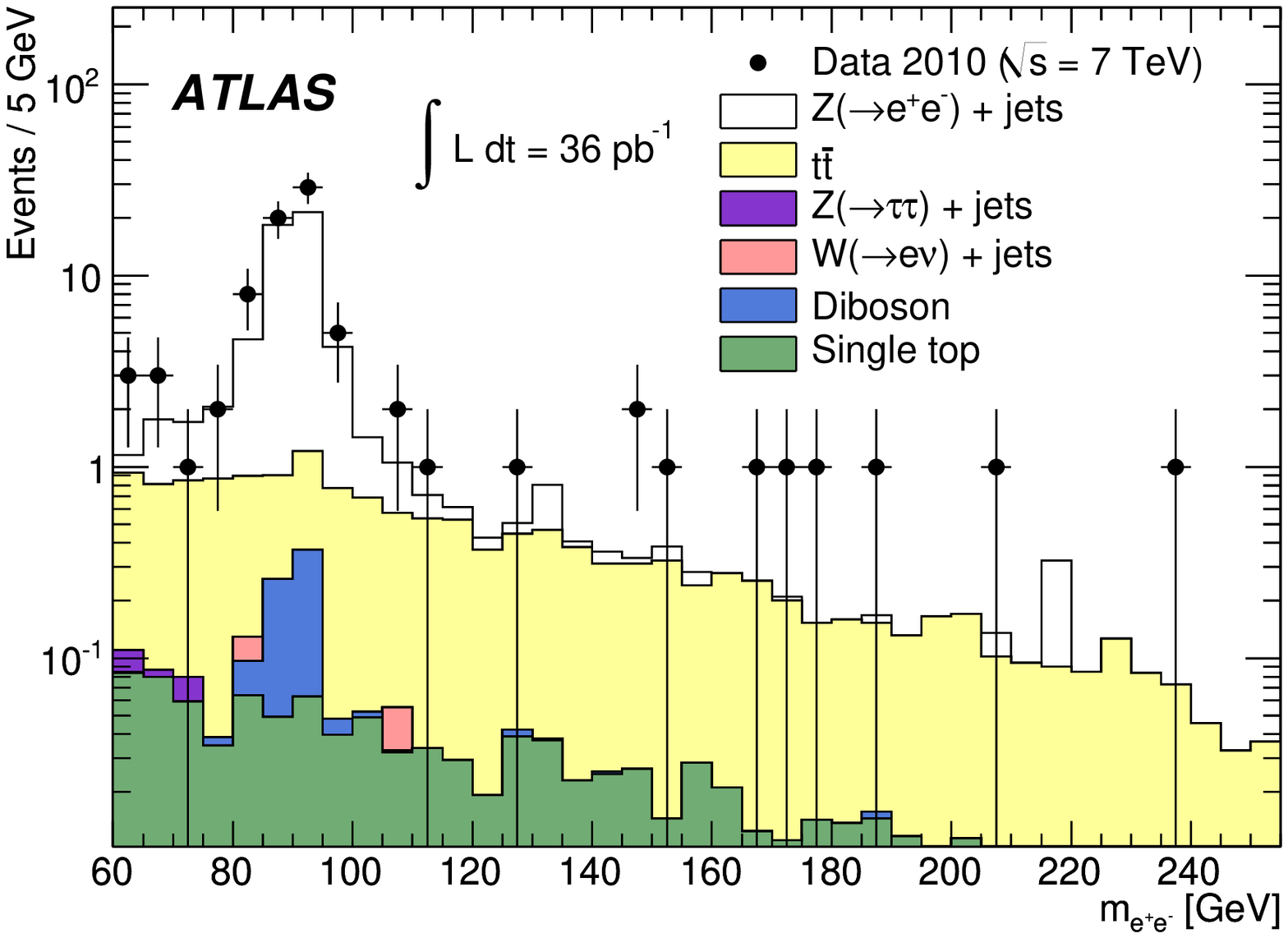}}
\subfigure{\includegraphics[width=0.49\textwidth]{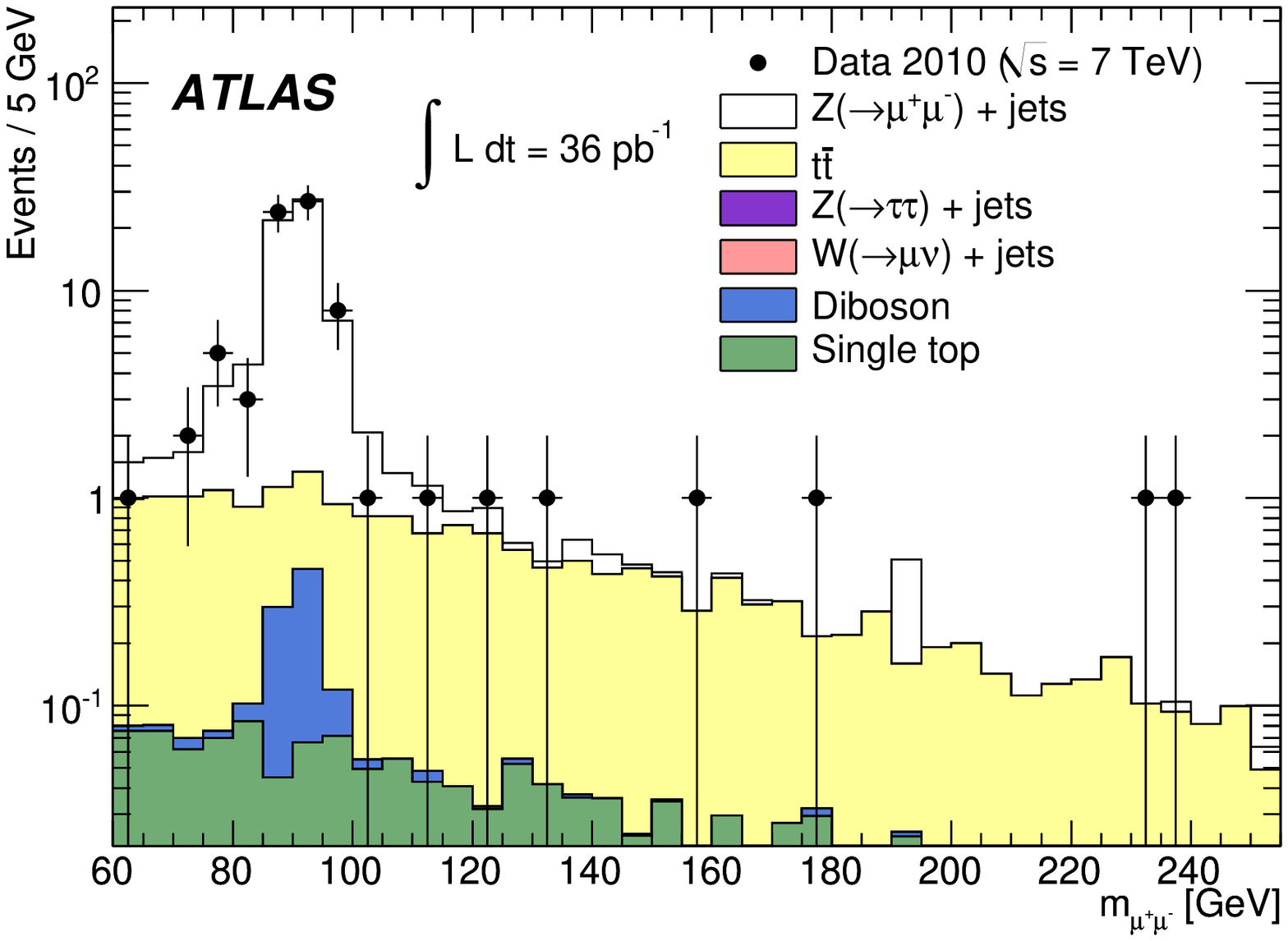}}
\caption{Top frames: Di-lepton mass distribution for events with at least one jet with $\pt>25~\GeV$ and
$|y|<2.1$, for the electron (left) and muon (right) channels. Bottom frames:  Di-lepton mass distribution for events with at 
least one $b$-tagged jet with $\pt>25~\GeV$ and $|y|<2.1$, for the electron (left) and muon (right) channels.
The contribution estimated from the simulated MC samples of the signal
and various background processes is shown. The small multi-jet background, 
estimated with a data-driven method, is not shown here (see text).}
\label{fig:llmass-jet}
\end{center}
\end{figure*}

The main background of $Z$ + $light$-jets and $Z$ + $c$-jets is taken
into account via the signal yield extraction procedure described below.
However, a number of other background processes
can contribute, namely $\ttbar$ with two leptonic decays of the $W$ bosons, 
$W$+ jets with one jet being mis-identified as an electron or muon, $Z$ + jets with the $Z$ boson decaying to $ \tau \tau$, dibosons $WW$/$WZ$/$ZZ$,
single-top production, and finally multi-jet production where the jets
contain real leptons from heavy flavour decays or are mis-identified
as leptons. The background processes other than that from multi-jets are
estimated from the MC simulation samples described in
Section~\ref{simevents}, resulting in the following contributions to
each channel after implementation of the analysis jet selection criteria:
 \ttbar\ ($\sim 6$ tagged jets), $WZ$ ($\sim 0.2$), $ZZ$ ($\sim 0.3$),
 single-top ($\sim 0.3$), and $Z \rightarrow \tau \tau $ ($\sim 0.1$).
In Fig.~\ref{fig:llmass-jet}, the top two frames show the di-lepton mass distribution for events with
 at least one jet with \pt~$>25$~GeV and
$|y|$~$<2.1$ for the electron and muon channels, together with the
simulation broken down into signal and background processes,
not including the multi-jet background. In the same figure, the bottom
two frames show the same distribution after requesting at least one jet that is $b$-tagged by the SV0 algorithm. 

The multi-jet background cannot be extracted reliably from simulation, and
hence is estimated using data-driven methods.
In the electron channel, the method considers two different samples of electron candidates with relaxed
selection criteria. In the first sample the selection criteria on one of the electrons are significantly relaxed, while
in the second sample, the criteria on both electrons are mildly relaxed but \it{Medium}\rm\ electrons are vetoed.
In these samples, the di-electron mass spectrum is fitted with two
components: the contribution of the signal and other background processes
modelled by the simulated MC samples (with relative signal/background
normalisation fixed by the MC predictions), and an exponential function
which was found to model the multi-jet contribution well. The fit parameters are the
normalisation of the MC component, and the 
normalisation and exponent of the multi-jet exponential. To determine the
multi-jet background the fit is repeated, but this time using the di-electron spectrum of events passing the full
analysis selection, and again allowing the normalisation of the MC
component and multi-jet background to float, but fixing the exponent of the
multi-jet exponential to the value determined from the relaxed selection samples.
The extracted exponents are similar for the two samples of relaxed electron candidates, leading
to the same estimated contribution of multi-jet events: $(1.0 \pm 2.2)$ events in the mass window
$76<m_{ll}<106$ GeV.
In the muon channel the method considers a sample of events similar to the signal but with non-isolated muon 
candidates, which has been shown to be dominated by the multi-jet background. The multi-jet background contribution to the signal
is then estimated assuming that the ratio of the number of events with isolated muons to that with non-isolated muons 
is the same in the data as in the simulated multi-jet MC sample, and amounts to $(0.0 \pm 0.9)$ events.

\subsection{Signal yield}

\begin{figure}[htp]
\begin{center}
\subfigure{\includegraphics[width=0.39\textwidth]{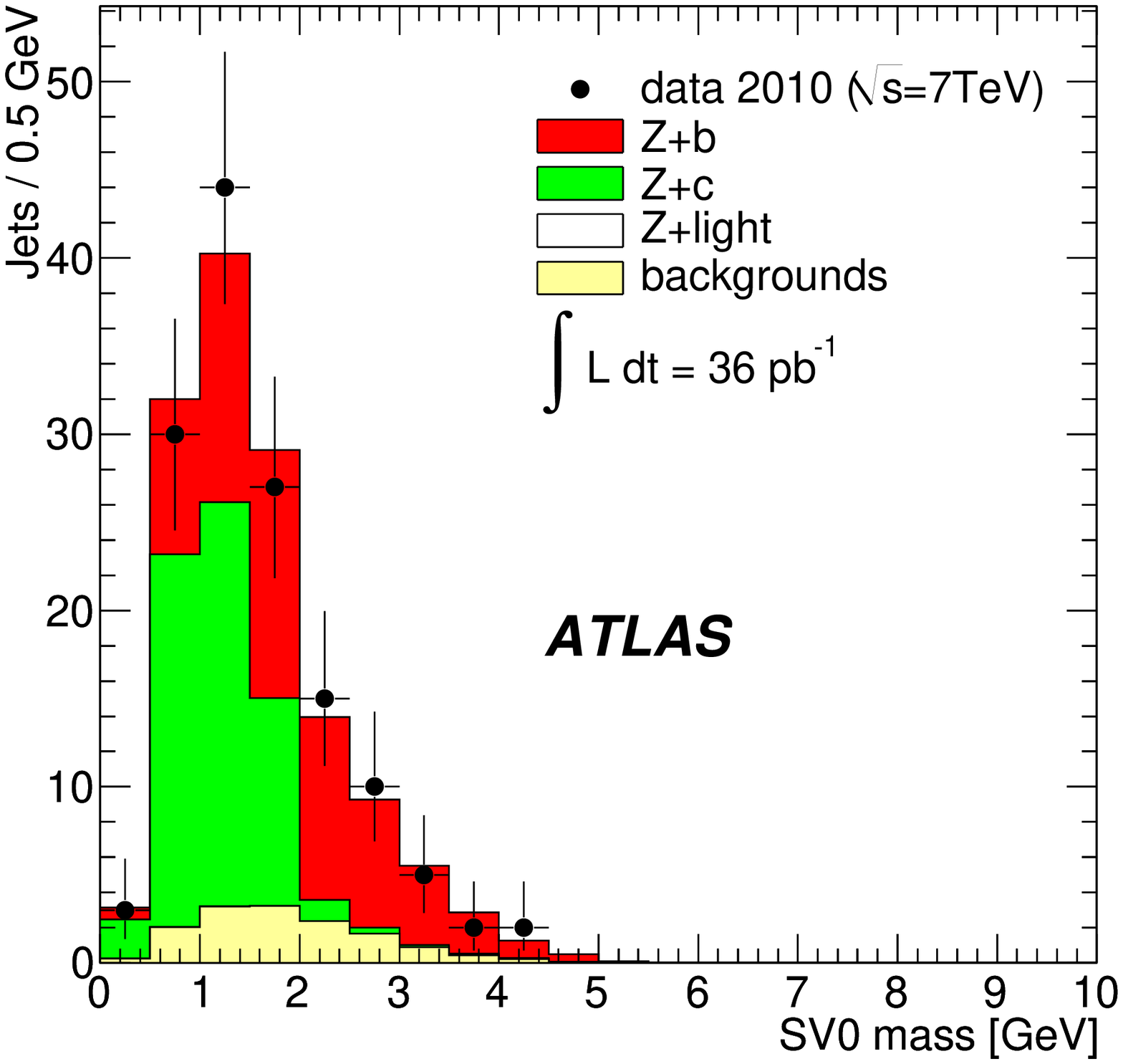}}
\caption{ SV0-mass distribution (see text) for $b$-tagged jets in the selected events.
The fitted contributions from $b$, $light$, and $c$-jets are displayed; the other background processes are also shown.}

\label{fig:sv0mass}
\end{center}
\end{figure}

\begin{table}[t]
\begin{center}
\begin{tabular}{lc}
\hline\hline
		&				\\
    $b$-jets 	&  $63.6^{+14.7}_{-13.2}$	\\
		&				\\
  $c$-jets 	&  $59.9^{+13.4}_{-14.0}$	\\
		&				\\
  $light$ jets    &  $0.0 ^{+5.1}$ 	\\
		&				\\
  Other backgrounds (fixed)    &  $14.5$ 	\\
\hline\hline
\end{tabular}
\caption{Number of jets for the various flavors in the combined
  electron and muon channel, determined from the fit 
to the SV0-mass distribution. The statistical
errors on the fit results are given.}
\label{tab:sv0massnum}
\end{center}
\end{table}

The jets that are $b$-tagged by the SV0 algorithm still contain $light$ and $c$-jets. 
The yield of $b$-jets is thus calculated
on a statistical basis, by fitting the expected contributions to the
SV0-mass distribution.
For the electron and muon channels,
templates are obtained from the simulation for the SV0-mass spectrum
of each contribution: signal $Z$ + $b$-jets, $Z$ + $light$ jets,
$Z$ + $c$-jets, and all other background processes ($\ttbar$, multi-jet, single-top, $W$ + jets, $Z \rightarrow \tautau$ and diboson). 
As the templates for the electron and muon channels are compatible, these channels are treated together
and both types of events are entered into one SV0-mass distribution. This spectrum is subjected
to a likelihood fit, consisting of a sum containing the fixed
contributions of the other background processes
and a floating amount of $Z$ + $b$, $Z$ + $light$, and $Z$ + $c$-jets.
In order to validate the fit procedure, we performed fit closure and linearity tests using pseudo-experiments.
In each pseudo-experiment, SV0-mass distributions are constructed using the templates with the same number of events
as in the data and varying proportions of the $b$-jet template, then the fit procedure is performed.
The results demonstrated good linearity and no bias.
Figure \ref{fig:sv0mass} shows the SV0-mass spectrum of $b$-jet candidates
with the fitted contributions, and Table \ref{tab:sv0massnum} gives the corresponding number of jets. The purity of
the $b$-tagged sample is therefore found to be about 46\% with this selection.

\section{Cross-section measurement and comparison to theory}
\label{cross}
\subsection{Unfolding to the particle level}

The particle-level cross-section, $\sigma_{b}$, is defined as
follows. The fiducial restrictions on the $Z$ decay are defined as lepton $\pt>20\GeV$
and $\abseta<2.5$, and di-lepton mass pair within the range
$76<m_{ll}<106 \GeV$. In this definition, ``dressed'' leptons are 
used to reconstruct the $Z$ \cite{dressedleptons} : the four-vectors of all photons in a cone of $\Delta R <0.1$ around 
the lepton are added to the lepton four-vector. Jets are
reconstructed from stable particles (particles with lifetime in excess of 10~ps) using the anti-$k_{t}$ algorithm with resolution parameter of $0.4$, and include 
muons and neutrinos.  Jets are required to satisfy  $\pt>25~\GeV$ and $|y|<2.1$, and jets within $\Delta R <0.5$ of either of
the $Z$ decay leptons are removed.  
At particle-level, a jet is considered to be a $b$-jet if there is a 
$b$-hadron with $\pt>5\GeV$ within $\Delta R <0.3$ of that jet,
 and only weakly-decaying $b$-hadrons are considered.

The per lepton channel cross-section is obtained from the experimental measurements using the following formula:

$$ \sigma_{b}=\frac{N_b}{C_{e} {{\cal L}_{e}} + C_\mu
  {{\cal L}_{\mu}}}$$

\noindent
where $N_b$ is the number of $b$-jets extracted from the fit to the
SV0-mass distribution of $b$-tagged jets (given in
Table~\ref{tab:sv0massnum}), $C_e$ and $C_{\mu}$ are the
acceptance factors for the electron and muon channels
respectively, and ${\cal L}_e$ and ${\cal L}_{\mu}$ are the respective 
integrated luminosities for each channel.

The $C_e$ and $C_{\mu}$ acceptance factors are
determined from the simulated MC samples of the signal process, 
and correspond to the probability that a particle-level $b$-jet in a
$Z$ event as defined above passes all of the jet and $Z$ event selection criteria at
the detector-level. They thus include the efficiency for a signal
event to pass the triggers used, the efficiency to reconstruct
electrons or muons, and the efficiency of the SV0 algorithm to tag
$b$-jets. The efficiencies of the electron and muon high $\pt$ triggers have
been studied with data, and for signal events in the acceptance
defined above the trigger efficiency is found to be in excess of 99\% for both electron 
 and muon channels. In the determination of the lepton and $b$-tagging
 efficiencies, scale factors to account for mis-modelling by the
 simulation are applied. In the case of electron and muon reconstruction,
 these scale factors are determined using inclusive $Z$ and $W$ events in data \cite{WZbaseline,egamma2010}, and
 are found to be close to unity at the few percent level. For the SV0
 $b$-tagging efficiency, the scale factors have been determined as a
 function of jet \pt~and $\eta$ using data events where the jet contains a
 reconstructed muon to enrich the $b$ component \cite{btag-eff}, and
 are found to be consistent with unity.
The uncertainty related to these
scale factors is propagated into the final uncertainty on the cross-section.

\subsection{Systematic uncertainties}\label{syst}

The systematic uncertainties on the measured fiducial $b$-jet cross-section are
summarised in Table~\ref{tab:Systematics}. A systematic source can
impact the measurement in two ways: it can affect the SV0-mass
template shapes and hence the fitted $N_b$, and/or it can affect the calculated
acceptance factors ($C_e$ and/or $C_{\mu}$). Where a systematic
affects both the fit result and the acceptance the corresponding correlations have been taken into account.

The dominant source of systematic uncertainty comes from 
the dependence of the measurement on the modelling of
the signal process in the simulation. The
calculated acceptance factors and SV0-mass template shapes both depend
on the assumed $b$-jet \pt~spectrum. The sensitivity to
the \pt~spectrum modelling in simulation is assessed
by reweighting
the \pt~spectrum of the simulation until a satisfactory agreement
with data in the \pt~spectrum of $b$-tagged jets was
observed. The resulting uncertainty is considerably larger than any
differences observed between ALPGEN~and SHERPA~in the SV0-mass fit
results and acceptance factors. In addition, we ascribe a small uncertainty due to the modelling of
MPI in the signal simulation, assessed by artificially doubling the
contribution of MPI events in the acceptance calculation.

In terms of the
reconstruction of $b$-jets, the main systematics enter via the
uncertainty on the $b$-tagging efficiency scale factors, and
uncertainty on the jet energy scale
(JES). The uncertainty on 
the $b$-tagging scale factors is estimated from studies of inclusive $b$-jets and 
$b$ semi-leptonic decays \cite{btag-eff}. The JES was studied extensively for
inclusive jets using simulations,
single hadron test-beam data and $\gamma$-jet events \cite{jes}. Its uncertainty was demonstrated to be below $5\%$ in the $\pt$ range considered here.
In simulation, the JES for $b$-jets was found to differ from that of $light$ jets by at most $2.5\%$. The
present statistics do not allow us to calibrate this difference from the data, hence an additional $2.5\%$ is added in quadrature
to the JES uncertainty of $b$-jets. The jet energy resolution was also considered and the uncertainty taken as that 
derived from $light$ jets. One also has to consider that the detector
simulation may not perfectly model the response to the SV0-mass distribution for
$light$, $c$, and $b$-jets. This uncertainty is estimated using control
samples of inclusive jet events that are enriched in heavy and light
flavour jets, and used to derive reweighting functions that can be
applied to the SV0-mass templates to account for data-simulation
disagreements. 

The impact of uncertainties in the background estimation are
small. The \ttbar~background is estimated purely from simulation, and the normalisation of this
background is varied according to the uncertainty on the NNLO
\ttbar~cross-section. The multi-jet backgrounds in both the electron and
muon channel are varied according to the uncertainties on these
estimates described above.

Other smaller sources of systematic uncertainty considered include
uncertainties on lepton reconstruction: the efficiency to reconstruct,
and the energy/momentum scale and resolution.  The estimation follows
closely that of the inclusive $Z$ analysis \cite{WZbaseline}, 
with the same methods applied to simulated $Z + b$ event samples. The uncertainties on the electromagnetic energy scale,
and on the muon momentum scale and resolution, all have a negligible
($<1\%$) impact.

\begin{table}[!ht]
  \centering
  
  \begin{tabular}{lcc}
  \hline
  \hline
			&	&	\\
 Source                 & SV0-mass Fit (\%) &  Acceptance (\%)	\\
			&	&	\\
\hline
\hline
\multicolumn{3}{c}{Both Electron and Muon}  	\\
\hline

$b$-tagging efficiency	& 1.7	& 9.1	\\
SV0-mass templates	& 3.5	&  -	\\
Model dependence	& 2.7	& 10.0 	\\
Jet energy scale	& 0.7   & 4.0	\\
$t\bar{t}$ cross-section& 2.0	&  - 	\\
MPI model               & negl. 	& 1.0	\\
\hline
\hline
\multicolumn{3}{c}{Electron only}  	\\
\hline
MC statistics 		& negl.	& 1.3  	\\
Multi-jet background 			& 1.6 	&  -	\\
Electron efficiency	& negl.	& 5.0	\\
 \hline
Total Electron          & 5.6	& 15.0 	\\
\hline
\hline
\multicolumn{3}{c}{Muon only}  \\
\hline
MC statistics 		& negl. 	& 1.3  	\\
Multi-jet background			& 0.7 	&  - 	\\
Muon efficiency 	& negl.	& 2.0	\\
\hline
Total Muon		& 5.4 	& 14.3	\\
\hline
\hline
			&	&	\\
Total Systematic	& \multicolumn{2}{c}{+21\% -16\%} \\
 Uncertainty	&	&	\\
  \hline
  \hline
\end{tabular}
 \caption{Fractional systematic uncertainties on the SV0-mass
   fit and acceptance results from each systematic source considered. Sources for which the shift is labelled ``negl.'' produced effects of less
   than 1\% which are negligible when added in quadrature (and not considered). When relevant,
   asymmetric errors are used for the calculation of the total, but only the average error is 
   shown for better readability. The ``Total Systematic Uncertainty'' result is the total percentage error 
   on the combined channel $b$-jet cross-section, and takes into account the correlations
   between SV0-mass fit and acceptance systematics.}
 \label{tab:Systematics}
 \end{table}%

\subsection{Results and comparison to theory}
The measured cross-section for $b$-jets produced in association with a $Z$ boson decaying into
one of the lepton channels is presented in Table \ref{tab:resultZb}, alongside values evaluated
in the different models presented in the introduction. The MCFM NLO
prediction is shown for the CTEQ6.6 PDF,  with the renormalisation and
factorisation scales taken as $\sqrt{M^2_Z + p^2_{T,Z}}$. 
In contrast to ALPGEN and SHERPA, MCFM does not simulate QED final state radiation (FSR)
nor non-perturbative hadronic effects. Correction factors are computed for lepton FSR, parton/jet 
correspondence, underlying event and MPI contribution, using events from particle-level LO simulations. 
The correction factor for non-perturbative hadronic effects is obtained by comparing particle-level 
results to parton-level, where parton-level jets are matched to $b$ quarks.
 This is calculated using SHERPA, PYTHIA and AcerMC~\cite{AcerMC}, with the spread of these results
defining the range of the correction, which is found to be $0.89 \pm 0.07$. 
The correction is dominated by the impact of $b$-hadron decay products falling outside the jet 
at the particle-level.
The correction factor for lepton FSR is similarly found to be $0.972 \pm 0.002$ for both lepton types,
 dominated by dilepton pairs migrating out of the required mass window.

In order to estimate theoretical uncertainties on the prediction, the renormalisation and factorisation scales
are independently shifted up, then down, by a factor of 2. The uncertainties arising from 
the different PDF error sets are also assessed, as well as using the CTEQ6.6 PDF with different
values of $\alpha_{S}$.
The raw MCFM prediction for the $Z + b$ cross-section in the fiducial region is
4.48 $^{+0.55} _{-0.56}$ (scale) $^{+0.10}_{-0.12}$ (PDF) $^{+0.08}_{-0.08}$ ($\alpha_s$) pb.
For comparison, the prediction obtained using the MSTW2008 PDF~\cite{mstw} is 
$4.80^{+0.62}_{-0.61}$ (scale) $^{+0.09}_{-0.10}$ (PDF)  $^{+0.09}_{-0.11} (\alpha_s)$ pb. 
The CTEQ6.6 and MSTW2008 PDFs use different default values for
$\alpha_s$ (propagated consistently through the NLO calculation), and, taking into account their combined PDF and $\alpha_s$ uncertainties, there is a marginal disagreement between the two predictions. 
However, given the precision of the experimental measurement, we cannot conclude that one PDF better
 reproduces the experimental result than the other. 
We quote the prediction using the CTEQ6.6 PDF by default, and the uncertainty quoted in Table \ref{tab:resultZb} for the corrected result corresponds to the quadratic sum of the uncertainties on the scale, PDF, $\alpha_s$, and the uncertainty on the non-perturbative correction.
The ALPGEN and SHERPA predictions are also shown, with errors from the
MC statistics only.

\begin{table}[t]
\begin{center}
\begin{tabular}{lc}
\hline
\hline
		&			\\
Experiment	&   $3.55^{+0.82}_{-0.74} (\mathrm{stat})^{+0.73}_{-0.55} (\mathrm{syst})\pm 0.12 (\mathrm{lumi})$ pb	\\
		&			\\
\hline
		&			\\	
MCFM		&   $3.88  \pm 0.58$ pb            \\
		&			\\	
ALPGEN		&   $2.23 \pm  0.01~(\mathrm{stat\ only})$ pb	\\
SHERPA		&   $3.29 \pm  0.04~(\mathrm{stat\ only})$ pb	\\
\hline\hline
\end{tabular}
\caption{Experimental measurement and predictions of $\sigma_b$, the
cross-section for inclusive $b$-jet production in association with a
$Z$ boson, per lepton channel, as defined in the text.}
\label{tab:resultZb}
\end{center}
\end{table}

\subsection{Measurement of the average number of $b$-jets per $Z$ event}

Simulation packages such as ALPGEN and SHERPA are based on LO
calculations 
and thus are not expected to accurately predict an absolute cross-section for the process studied here. 
However, they are often used to
generate fully simulated events for the study of backgrounds to the search of other processes, as mentioned 
in the introduction. A current practice is then to normalise the cross-section of generated events to that of 
a well known, more inclusive process. In this approach, the analysis presented here is extended 
to measure the ratio of $\sigma_b$  to that of
 the cross-section for the inclusive production of the $Z$ boson (for the same fiducial
 restrictions on the $Z$ decay), i.e. the average number of $b$-jets per $Z$ event.
 To obtain the inclusive $Z$ sample, the analysis is repeated with the same selection as above, except
the jet requirements. The cross-section obtained for the inclusive $Z$ production with the same fiducial 
region for the leptons is $465 \pm 3$ pb (statistical error only), in agreement with the ATLAS 
measurement \cite{WZ2010}. The systematic uncertainties on the ratio are propagated coherently in 
the $Z + b$ and $Z$ selections. The uncertainties related to leptons cancel to a negligible level, 
and those related to luminosity cancel completely. However as the main systematic uncertainties concern only the
 $Z + b$ analysis ($b$-tagging, model dependence, jet energy scale), the overall systematic uncertainty
 is only marginally reduced.
  
 The MCFM prediction of this ratio is calculated with the same method and assumptions as above. 
 To estimate the systematic uncertainty, the scale and PDF choices are varied coherently between 
the $Z+b$ and inclusive $Z$ samples for each sub-process simulated. 
Table \ref{tab:resultZbZ} shows the experimentally measured result for
the average number of $b$-jets per $Z$ event and 
comparisons to the theoretical predictions. The MCFM NLO prediction is in agreement with the data.
The ALPGEN and SHERPA predictions differ significantly from each other,
but are both compatible with the data within the experimental uncertainties.
 \begin{table}[t]
\begin{center}
\begin{tabular}{lc}
\hline
\hline
		&			\\
Experiment	&  $(7.6^{+1.8}_{-1.6} (\mathrm{stat})^{+1.5}_{-1.2} (\mathrm{syst}))\times 10^{-3}$	\\
		&			\\
\hline
		&			\\	
MCFM		&   $(8.8 \pm 1.1)\times 10^{-3}$             \\
		&			\\	
ALPGEN		&   $(6.2 \pm 0.1~(\mathrm{stat\ only}))\times 10^{-3}$ 	\\
SHERPA		&   $(9.3 \pm 0.1~(\mathrm{stat\ only}))\times 10^{-3}$ 	\\
\hline\hline
\end{tabular}
\caption{Experimental measurement and predictions of the average
  number of $b$-jets produced in association with a $Z$ boson, with
  the same fiducial region as defined in the text for $\sigma_b$.}
\label{tab:resultZbZ}
\end{center}
\end{table}

\section{Conclusions}
\label{conclusion}
A first measurement is made of the cross-section for the production of
$b$-jets in association with a $Z$ boson in proton-proton collisions at $\sqrt{s} = 7$ TeV,
using ~$36~\ipb$ of data collected in 2010 by the ATLAS experiment. In addition, the average
number of $b$-jets per $Z$ event is extracted. 
Both measurements are currently statistics limited.
The predictions from NLO pQCD calculations agree well with both
results. Leading order generators are able to reproduce the measured average number of
$b$-jets per $Z$ event within the uncertainties of the measurement,
although their predictions differ significantly from each other.

\section*{Acknowledgements}

We thank CERN for the very successful operation of the LHC, as well as the
support staff from our institutions without whom ATLAS could not be
operated efficiently.

We acknowledge the support of ANPCyT, Argentina; YerPhI, Armenia; ARC,
Australia; BMWF, Austria; ANAS, Azerbaijan; SSTC, Belarus; CNPq and FAPESP,
Brazil; NSERC, NRC and CFI, Canada; CERN; CONICYT, Chile; CAS, MOST and
NSFC, China; COLCIENCIAS, Colombia; MSMT CR, MPO CR and VSC CR, Czech
Republic; DNRF, DNSRC and Lundbeck Foundation, Denmark; ARTEMIS, European
Union; IN2P3-CNRS, CEA-DSM/IRFU, France; GNAS, Georgia; BMBF, DFG, HGF, MPG
and AvH Foundation, Germany; GSRT, Greece; ISF, MINERVA, GIF, DIP and
Benoziyo Center, Israel; INFN, Italy; MEXT and JSPS, Japan; CNRST, Morocco;
FOM and NWO, Netherlands; RCN, Norway; MNiSW, Poland; GRICES and FCT,
Portugal; MERYS (MECTS), Romania; MES of Russia and ROSATOM, Russian
Federation; JINR; MSTD, Serbia; MSSR, Slovakia; ARRS and MVZT, Slovenia;
DST/NRF, South Africa; MICINN, Spain; SRC and Wallenberg Foundation,
Sweden; SER, SNSF and Cantons of Bern and Geneva, Switzerland; NSC, Taiwan;
TAEK, Turkey; STFC, the Royal Society and Leverhulme Trust, United Kingdom;
DOE and NSF, United States of America.

The crucial computing support from all WLCG partners is acknowledged
gratefully, in particular from CERN and the ATLAS Tier-1 facilities at
TRIUMF (Canada), NDGF (Denmark, Norway, Sweden), CC-IN2P3 (France),
KIT/GridKA (Germany), INFN-CNAF (Italy), NL-T1 (Netherlands), PIC (Spain),
ASGC (Taiwan), RAL (UK) and BNL (USA) and in the Tier-2 facilities
worldwide.

\bibliographystyle{model1-num-names}
\bibliography{Zb}
\onecolumn
\clearpage
\input{atlas_authlist.tex}

\end{document}

%% file: atlas_authlist.tex
\begin{flushleft}
{\Large The ATLAS Collaboration}

\bigskip

G.~Aad$^{\rm 48}$,
B.~Abbott$^{\rm 111}$,
J.~Abdallah$^{\rm 11}$,
A.A.~Abdelalim$^{\rm 49}$,
A.~Abdesselam$^{\rm 118}$,
O.~Abdinov$^{\rm 10}$,
B.~Abi$^{\rm 112}$,
M.~Abolins$^{\rm 88}$,
H.~Abramowicz$^{\rm 153}$,
H.~Abreu$^{\rm 115}$
% ...
\bigskip

% ...
$^{10}$ Institute of Physics, Azerbaijan Academy of Sciences, Baku, Azerbaijan\\
$^{11}$ Institut de F\'isica d'Altes Energies and Departament de F\'isica de la Universitat Aut\`onoma  de Barcelona and ICREA, Barcelona, Spain\\
% ...
$^{48}$ Fakult\"{a}t f\"{u}r Mathematik und Physik, Albert-Ludwigs-Universit\"{a}t, Freiburg i.Br., Germany\\
$^{49}$ Section de Physique, Universit\'e de Gen\`eve, Geneva, Switzerland\\
% ...
$^{88}$ Department of Physics and Astronomy, Michigan State University, East Lansing MI, United States of America\\
% ...
$^{111}$ Homer L. Dodge Department of Physics and Astronomy, University of Oklahoma, Norman OK, United States of America\\
$^{112}$ Department of Physics, Oklahoma State University, Stillwater OK, United States of America\\
% ...
$^{115}$ LAL, Univ. Paris-Sud and CNRS/IN2P3, Orsay, France\\
% ...
$^{118}$ Department of Physics, Oxford University, Oxford, United Kingdom\\
% ...
$^{153}$ Raymond and Beverly Sackler School of Physics and Astronomy, Tel Aviv University, Tel Aviv, Israel\\
% ...
\end{flushleft}